\documentstyle[12pt,aps,prd]{revtex}
\input epsf

\newcommand{\beq}{\begin{equation}}
\newcommand{\eeq}{\end{equation}}

\def\lap{\lower.5ex\hbox{$\; \buildrel < \over \sim \;$}}
\def\gap{\lower.5ex\hbox{$\; \buildrel > \over \sim \;$}}
\def\lesssim{\lap}
\def\gtrsim{\gap}

\def\L{\Lambda}

\begin{document}

\title{Anthropic prediction for $\Lambda$ and the $Q$ catastrophe}
\author{Jaume Garriga $^1$ and Alexander Vilenkin $^2$}
\address{
$^1$ Departament de F{\'\i}sica Fonamental, Universitat de Barcelona,\\
Mart{\'\i}\ i Franqu{\`e}s 1, 08193 Barcelona, Spain\\
$^2$ Institute of Cosmology, Department of Physics and Astronomy,\\
Tufts University, Medford, MA 02155, USA\\
}

\maketitle

\begin{abstract}

We discuss probability distributions for the cosmological constant
$\Lambda$ and the amplitude of primordial density fluctuations $Q$ in
models where they both are anthropic variables. With mild assumptions
about the prior probabilities, the distribution $P(\Lambda,Q)$
factorizes into two independent distributions for the variables $Q$
and $y \propto \Lambda/Q^3$. The distribution for $y$ is largely
model-independent and is in a good agreement with the observed value
of $y$. The form of $P(Q)$ depends on the origin of density
perturbations. If the perturbations are due to quantum fluctuations of
the inflaton, then $P(Q)$ tends to have an exponential dependence on
$Q$, due to the fact that in such models $Q$ is correlated with the
amount of inflationary expansion. For simple models with a power-law
potential, $P(Q)$ is peaked at very small values of $Q$, far smaller
than the observed value of $10^{-5}$. This problem does not arise in
curvaton-type models, where the inflationary expansion factor is not
correlated with $Q$. 

\end{abstract}

\section{Introduction}

The fundamental theory of nature may admit multiple vacua with
different low-energy constants. This possibility has attracted much
attention in the context of ``string theory landscape''
\cite{Bousso,Susskind,Douglas}.  When combined with the theory of
eternal inflation \cite{AV83,Linde86}, it leads to the picture of a
``multiverse'', where constants of nature take different values in
different post-inflationary (thermalized) regions of spacetime. The
key problem in this theoretical framework is to calculate the
probability distribution for the constants. Once we have the
distribution, we can use the principle of mediocrity - the assumption
that we are typical among the observers in the universe - and make
predictions for the constants at a specified confidence level.

A major success of this program has been the prediction of a non-zero
cosmological constant $\Lambda$ \cite{Weinberg87,AV95,Efstathiou,MSW}. 
The observed value of $\Lambda$ is well within the $2\sigma$ range of the
theoretical distribution. Recently, however, it has
been argued that this successful prediction does not survive when
other parameters, such as the amplitude of primordial density
fluctuations $Q$, are also allowed to vary \cite{Banks,Wise}. Here, we
critically examine these claims and discuss some issues surrounding the
calculation of prior probabilities. We start with a review
of the original prediction for $\Lambda$.

After this work was completed, we noticed the paper \cite{FHW} by
Feldstein, Hall and Watari, which has a substantial overlap with our
work. 

\section{Anthropic prediction for $\Lambda$}

The key observation, due to Weinberg, is that
structure formation effectively stops when $\Lambda$ comes to dominate
the universe. If no galaxies are formed by that time, then in the
future there will also be no galaxies, and therefore no observers. 
The corresponding anthropic bound on the cosmological constant is
\cite{Weinberg87} 
\beq
\Lambda\lesssim 500\rho_{m0}
\label{bound}
\eeq
where $\rho_{m0}$ is the present matter density. A lower bound on
$\Lambda$ can be obtained by requiring that the universe does not
recollapse before life had a chance to develop. Assuming that the
timescale for life evolution is comparable to the present cosmic time,
one finds \cite{Barrow,LK} $\Lambda\gtrsim -\rho_{m0}$.

The anthropic bound (\ref{bound}) specifies the value of $\Lambda$
which makes galaxy formation barely possible. However, if the
effective $\Lambda$ varies in space, then most of the galaxies will
not be in regions characterized by this marginal value, but rather in
regions teeming with galaxies, where $\Lambda$ dominates after a
subatantial fraction of matter has already clustered.

To make this quantitative, we define the probability distribution
${P}(\L)d\L$ as being proportional to the number of observers
in the universe who will measure $\L$ in the interval $d\L$. This
distribution can be represented as a product \cite{AV95}
\beq
{P}(\L)d\L=n_{obs}(\L){P}_{prior}(\L) d\L.
\label{dP}
\eeq
Here, ${P}_{prior}(\L)d\L$ is the prior distribution, which is
proportional to the volume of those parts of the universe where
$\L$ takes values in the interval $d\L$, and $n_{obs}(\L)$ is
the number of observers that are going to evolve per unit volume.
The distribution (\ref{dP}) gives the probability that a
randomly selected observer is located in a region where the effective
cosmological constant is in the interval $d\L$.  

Of course, we have no idea how to calculate $n_{obs}$, but what comes
to the rescue is the fact that the value of $\L$ does not directly
affect the physics and chemistry of life. As a rough approximation, we
can then assume that $n_{obs}(\L)$ is simply proportional to the
fraction of matter $f$ clustered in giant galaxies like ours
(with mass $M\gtrsim M_G = 10^{12}M_{\odot}$),
\beq
n_{obs}(\L)\propto f_G(\L).
\label{nobs}
\eeq
The idea is that there is a certain number of stars per unit mass in a
galaxy and certain number of observers per star. The same
approximation can be used for other ``life-neutral'' parameters, like
the amplitude of primordial density fluctuations $Q$ or the neutrino
masses.

Now we have to decide what should be used for the prior probability in
(\ref{dP}). At present, the details of the fundamental theory and of
the inflationary dynamics are too uncertain for a definitive
calculation of the prior. We shall instead rely on the following
general argument \cite{AV96,Weinberg96}.

Suppose some parameter $X$ varies in the range $\Delta X$ and is
characterized by a prior distribution ${P}_{prior}(X)$. Suppose
further that $X$ affects the number of observers in such a way that
this number is non-negligible only in a very narrow range $\Delta
X_{obs}\ll \Delta X$. Then one can expect that the function
${P}_{prior}(X)$ with a large characteristic range of variation should
be very nearly a constant in the tiny interval $\Delta
X_{obs}$. In the case
of $\L$, the range $\Delta\L$ is set by the Planck scale or by
the supersymmetry breaking scale, and we have $(\Delta
\L)_{obs}/\Delta\L \sim 10^{-60} -10^{-120}$. Hence, we 
expect
\beq
{P}_{prior}(\L)\approx {\rm const}.
\label{prior}
\eeq
We emphasize that the assumption here is that the value $\L=0$ is not
in any way special, as far as the fundamental theory is concerned, and
is, therefore, not a singular point of ${P}_{prior}(\L)$.

Combining Eqs.(\ref{dP}),(\ref{nobs}),(\ref{prior}), we obtain
\beq
{P}(\L)\propto f_G(\L).
\label{Pf}
\eeq
The fraction of clustered matter, $f_G(\L)$, can be calculated using
the Press-Schechter approximation. Restricting attention to positive
values of $\Lambda$, one finds \cite{MSW,GV03}
\beq
f_G(\L)\propto {\rm erfc}\left[0.8 \left({\L\over{\rho_m \sigma_G^3}}
\right)^{1/3} \right],
\label{PS}
\eeq
where $\sigma_G$ is the density contrast on the galactic scale and
$\rho_m$ is the matter density. The
product $\rho_m\sigma_G^3$ is time-independent during the matter era,
so it can be evaluated at any time.

Introducing a dimensionless variable
\beq
y\equiv \L/\rho_m\sigma_G^3, 
\label{ydef}
\eeq
we can write the distribution as
\beq
dP(y)\propto {\rm erfc}(0.8 y^{1/3})y d(\ln y).
\label{Py}
\eeq
This distribution is peaked at $y \sim 1$. Discarding 2.5\% at both
tails of (\ref{Py}) yields the range 
\beq
0.043 < y < 16. 
\label{y}
\eeq
This is the
prediction for $y$ at 95\% confidence level.

The product $\rho_m\sigma_G^3$ can be expressed in terms of the
present-day values, $\rho_{m0}$ and $\sigma_{G0}$. Hence, $y$ is a
measurable quantity, and the anthropic prediction can be tested. As
already mentioned, the observed value of $y$ is well within the 95\%
range (\ref{y}) (see, e.g., \cite{GLindeV}).

Apart from this successful prediction, two additional factors make the
anthropic explanation of the observed value of $\L$ particularly
compelling. First, there are no plausible alternatives. Second, the
anthropic approach also gives a natural resolution to the coincidence
puzzle: Why do we happen to live at the very special epoch when
$\Lambda \sim \rho_m$? If we denote the redshifts at the epochs of
galaxy formation and of $\L$-domination by $z_G$ and $z_\L$,
respectively, then most of the galaxies should be in regions where
$z_\L \lesssim z_G$. Regions with $z_\L\gg z_G$ will have very few
galaxies, while regions with $z_\L\ll z_G$ will be rare simply because
they correspond to a very narrow range of $\L$ near zero (small ``phase
space''). Hence, we expect a typical galaxy to be located in a region
where
\beq
z_\L \sim z_G.
\label{coincidence}
\eeq
The galaxy formation epoch, $z_G\sim 1-3$, is close to the present
cosmic time. This explains the coincidence \cite{GLV,Bludman}.

\section{Variable $Q$}

Banks {\it et. al.} \cite{Banks} and Graesser {\it et. al.}
\cite{Wise} have argued that the successful prediction for $\Lambda$
will be destroyed if the density fluctuation amplitude $Q$ is also
allowed to vary. $Q$ is defined as the density contrast at the time of
horizon crossing. It is approximately the same on all scales of
astrophysical interest. The anthropically allowed range of $Q$ is
\cite{TR} 
\beq
10^{-6} \lesssim Q \lesssim 10^{-4}. 
\label{Qrange}
\eeq
The observed value,
$Q\sim 10^{-5}$, is in the middle of this range.

For larger values of $Q$, galaxies form earlier, so $\L$ can get
larger without interfering with the galaxy formation process,
$\L_{max}\propto Q^3$. Since larger $\L$ correspond to larger phase space,
this suggests that the probability is maximized for large $Q$ and
$\L$, e.g., $Q\sim 10^{-4}$, $\L\sim 10^3 \rho_{m0}$. 

To examine the situation more carefully, let us now consider the
probability distribution for $\L$ and $Q$. Since the galactic-scale
density contrast $\sigma_G$ is linearly related to $Q$,
$\sigma_G\propto Q$, we can equivalently consider the distribution for
$\L$ and $\sigma_G$, 
\beq
dP(\L,\sigma_G)\propto P_{prior}(\L,\sigma_G){\rm erfc}(0.8 y^{1/3})
d\L d\sigma_G,
\label{PLs}
\eeq
where, as before, $y$ is given by Eq.~(\ref{ydef}). The narrow anthropic range
of $\L$ suggests, also as before, that the prior probability is
independent of $\L$ in the range of interest,
\beq
P_{prior}(\L,\sigma_G)\approx P_{prior}(\sigma_G).
\eeq
Substituting this in (\ref{PLs}) and changing variables from
$(\L,\sigma_G)$ to $(y,\sigma_G)$, we obtain \cite{GLV}
\beq
dP(y,\sigma_G)\propto \sigma_G^3 P_{prior}(\sigma_G) d\sigma_G \cdot
{\rm erfc}(0.8 y^{1/3}) dy.
\label{Pys}
\eeq

Remarkably, the distribution for $y$ and $\sigma_G$ is
factorized. Moreover, the $y$-part of the distribution is exactly the
same as it was for the original model, where $\sigma_G$ was not
variable. This means that {\it the successful prediction for $y$ is
completely unaffected by the variation of $\sigma_G$} \cite{GV03}.

\section{The large $Q$ catastrophe}

Let us now consider the distribution for $\sigma_G$,
\beq
dP(\sigma_G)\propto \sigma_G^3 P_{prior}(\sigma_G) d\sigma_G.
\label{Ps}
\eeq
The value of $\sigma_G$ is determined by the horizon-crossing
perturbation amplitude $Q$, which is in turn determined by the
inflaton potential $V(\phi)$,
\beq
\sigma_G\propto Q\propto {V^{3/2}\over{V'}},
\label{sQ}
\eeq
where the right-hand side is evaluated at the value of $\phi$
corresponding to horizon crossing for the comoving scale of interest.
With a power-law potential
\beq
V(\phi)=\lambda\phi^n,
\label{V}
\eeq 
this gives, neglecting logarithmic factors (see, e.g.,
\cite{Lindebook}), 
\beq 
Q\propto\lambda^{1/2}.
\label{slambda}
\eeq

Suppose now that $\lambda$ is a variable, which may be determined by some
additional scalar field. It seems natural to assume that the range of
variation for $\lambda$ is $\Delta\lambda\sim 1$.  The observed value
of $\sigma_G$ is obtained for $\lambda \sim 10^{-14}$, and the
anthropic range (\ref{Qrange}) corresponds to $10^{-16} \lesssim
\lambda \lesssim 10^{-12}$. Since this range is so narrow, Graesser
{\it et. al.} \cite{Wise} argue that the same logic we used for the
cosmological constant implies that the prior for $\lambda$ should be
nearly flat in the range of interest, 
\beq 
dP_{prior}/d\lambda\approx {\rm const}.
\label{flatlambda}
\eeq
Then it follows from (\ref{slambda}) that $dP_{prior}(Q)\propto
Q dQ$, and Eq.~(\ref{Ps}) gives
\beq
dP(Q)\propto Q^4 dQ.
\label{s4}
\eeq

This distribution is strongly biased in favor of large values of
$Q$.  If anthropic factors cut off the distribution above
$Q^{(max)} \sim 10^{-4}$, then Eq.~(\ref{s4}) suggests that this
cutoff value is $10^5$ times more probable than the observed value
$Q\sim 10^{-5}$. This is the large $Q$ catastrophe.

We note, however, that there is an important difference between the
cosmological constant and the inflaton self-coupling $\lambda$, which
may invalidate the argument for the flat prior
(\ref{flatlambda}). Unlike the small cosmological constant, the value
of $\lambda$ has a strong effect on the dynamics of inflation. As we
shall see later, smaller values of $\lambda$ give larger inflationary
expansion factors. Hence, $\lambda=0$ is a rather special value of the
coupling, and the flat prior assumption is not justified.

\section{The problem of the prior}

The calculation of prior probabilities is a very challenging and
important problem. The main difficulty is that the volume of
thermalized regions with any given values of the parameters (such as
$Q$ or $\L$) is infinite, even for a region of a finite comoving
size. To compare such infinite volumes, one has to introduce some sort
of a cutoff. For example, one could include only regions that
thermalized prior to some time $t_{c}$ and evaluate volume ratios in
the limit $t_{c}\rightarrow\infty$. However, one finds that the
results are highly sensitive to the choice of the cutoff procedure (in
this case, to the choice of the time coordinate $t$~)\cite{LLM}. (For
a recent discussion, see \cite{Guth,Tegmark}.)

An eternally inflating universe consists of isolated thermalized
regions embedded into the inflating background of false vacuum.  These
thermalized islands are rapidly expanding into the inflating sea, but
the gaps between them are also expanding, making room for more
thermalized islands to form. The thermalization surfaces at the
boundaries between inflating and thermalized spacetime domains are
three-dimensional, infinite, spacelike hypersurfaces. The spacetime
geometry of an individual thermalized domain is most naturally described by
choosing the corresponding thermalization surface as the origin of
time. The domain then appears as a self-contained, infinite
universe, with the thermalization surface playing the role of the big
bang. Following Alan Guth, we shall call such infinite domains
``pocket universes''. All pocket universes are spacelike-separated and
are, therefore, causally disconnected from one another.\footnote{In
models where false vacuum decays through bubble nucleation, the role
of pocket universes is played by individual bubbles.} 

The reason for the cutoff-dependence of probabilities is that the
volume of an eternally inflating universe is growing exponentially
with time. The volumes of regions with all possible values of the
parameters are growing exponentially as well. At any time, a
substantial part of the total thermalized volume is ``new'' and thus
close to the cutoff surface. It is not surprising, therefore, that the
result depends on how that surface is drawn.

It has been suggested in \cite{AV98} that the resolution of this
difficulty may lie in the direction of switching from a global
distribution, defined with the aid of some globally defined time
coordinate, to a pocket-based distribution. In a particular case, when
the variable parameters take all their values within each pocket
universe, the problem appears to have been solved completely. In this
case, all pocket universes are statistically equivalent, and we can
pick any one of them to calculate the distribution. The prior
probability for the constants can be identified with the volume
fraction occupied by the corresponding regions of the pocket
universe. We can find this fraction by first evaluating it within a
sphere of large radius $R$ and then taking the limit
$R\to\infty$. This is the so-called ``spherical cutoff method''.

In general, however, there are several distinct types of pocket
universes, and we have to face the problem of comparing volumes in
different pockets. It is not clear how this can be done, since the
spherical cutoff method cannot be applied to disconnected spaces.  The
problem of defining probabilities in the general case still remains
unresolved. A conjecture toward this goal will be discussed in the
following section.

To get some idea of what the prior distribution for $Q$ may be like,
we shall first focus on the case of identical pockets, where we know
how to calculate probabilities. The spacetime structure of a pocket
universe is illustrated in Fig.~1. The surface $\Sigma_*$ in the
figure is the thermalization surface. It is the boundary between
inflating and thermalized domains of spacetime, which marks the end of
inflation and plays the role of the big bang in the pocket
universe. The surface $\Sigma_q$ is the boundary between the
stochastic domain, where the dynamics of the inflaton field is
dominated by quantum diffusion, and the deterministic domain, where it
is dominated by the deterministic slow roll. Thus, $\Sigma_q$ marks
the onset of the slow roll.

\begin{figure}
\begin{center}
\leavevmode\epsfxsize=5in\epsfbox{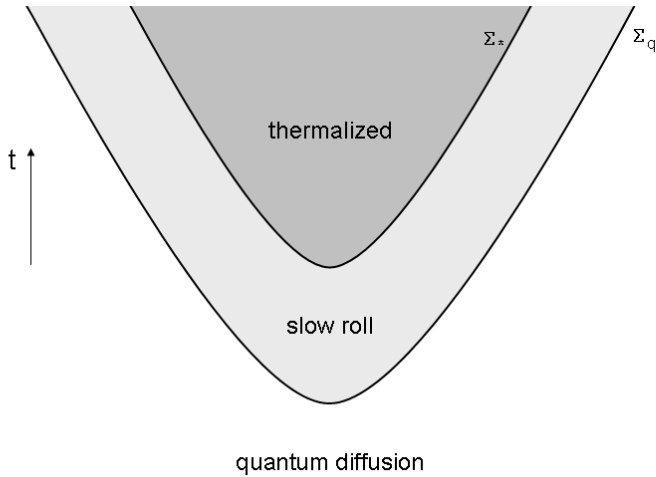}
\end{center}
\caption{The spacetime structure of a pocket universe.}
\end{figure}

The prior probability $P_{prior}(X)dX$ for a variable $X$ is defined in
terms of the volume fraction on the thermalization hypersurface
$\Sigma_*$.  It can be expressed as
\beq
P_{prior}(X)\propto P_q(X)Z^3(X),
\label{Pq}
\eeq
where $P_q(X)$ is the distribution (volume fraction) on
$\Sigma_q$, and $Z(X)$ is the volume expansion factor during the slow
roll. We assume for simplicity that the field responsible for the
value of $X$ interacts very weakly with the inflaton, so that $X$ does
not change appreciably during the slow roll. Otherwise, we would have
to include an additional Jacobian factor in Eq.~(\ref{Pq}), as
indicated in \cite{AV98}.

The expansion factor $Z(X)$ can be found from
\beq
Z(X)\approx \exp\left[4\pi\int_{\phi_{*X}}^{\phi_{qX}} {H(\phi,X)
\over{H'(\phi,X)}} d\phi\right],
\label{Z}
\eeq
where
\beq
H(\phi;X)=[8\pi V(\phi,X)/3]^{1/2}
\label{HV}
\eeq
is the inflationary expansion rate, $V(\phi,X)$ is the inflaton
potential, and prime stands for derivative with respect to
$\phi$. $\phi_{qX}$ and $\phi_{*X}$ are the values of $\phi$ at the
boundary surfaces $\Sigma_q$ and $\Sigma_*$, respectively. They are
defined by the conditions
\beq
H'/H^2(\phi_{qX},X)\sim 1
\label{phiq}
\eeq
and
\beq
H'/H(\phi_{*X},X)\sim 1.
\label{phi*}
\eeq

The distribution $P_q(X)$ can in principle be determined from
numerical simulations of the quantum diffusion regime. Some useful
techniques for this type of simulation have been developed in
\cite{LLM,VVW}. The analysis is greatly simplified in a class of models where 
the potential is independent of $X$ in the diffusion regime, and
$P_q(X)$ can be determined from symmetry,
\beq
P_q(X) = {\rm const}.
\label{Pqflat}
\eeq
An example is a ``new'' inflation type model with a complex inflaton,
$\phi = |\phi| \exp (iX)$. Inflation occurs near the maximum of the
potential at $\phi=0$, and we assume in addition that the potential is
symmetric near the top, $V = V(|\phi|)$. Eq.~(\ref{Pqflat}) follows if
this property holds throughout the diffusion regime. In such models,
the distribution (\ref{Pq}) reduces to
\beq
P_{prior}(X)\propto Z^3(X).
\label{PZ}
\eeq
Eq.~(\ref{PZ}) has a simple intuitive meaning: the prior probability
is determined by the volume expansion factor during the slow roll.

\section{A conjecture for a more general case}

In more general models, the factor $P_q(X)$ may provide an additional
dependence on $X$. It has been recently conjectured \cite {GPVW} that this
factor is of the form
\beq
P_q(X) \propto \exp [S(X)],
\label{ergodic}
\eeq
where 
\beq
S(X)=\pi/H^2(\phi_{qX},X)
\label{S}
\eeq
is the Gibbons-Hawking entropy of de Sitter space. 

The idea is that the evolution in the diffusion regime is {\it
ergodic}, in the sense that all quantum states are explored with equal
weight. The conjecture is that this ergodic property extends to the
diffusion boundary $\Sigma_q$. The right-hand side of
Eq.(\ref{ergodic}) is simply the number of quantum states for the
specified values of the parameters (fields) $X$ on the boundary. The
conjecture (\ref{ergodic}) has been verified in some specific models,
but the conditions of its validity are not presently clear.

In models with pockets of several different types, we need to
introduce an additional factor $p_j$, characterizing the relative
probability of different pockets. The full expression for the prior
probability is then given by 
\beq 
P_{prior}(X;j) \propto p_j e^{S(X;j)} Z^3(X;j),
\label{fullprior}
\eeq
where the index $j$ labels different types of pockets. 
A natural choice for the factor $p_j$ is the so-called comoving
probability, which can be defined as the probability for a test
particle, starting near the top of the potential $V(\phi,X)$, to end
up in a pocket of type $j$. This probability can be easily calculated
using the Fokker-Planck equation. 

One problem with this definition of $p_j$ is that it has some
dependence on the initial conditions assumed for the test particle. For
example, if the potential has several peaks of comparable height,
different values of $p_j$ will be obtained starting from different
peaks. An alternative definition of $p_j$, which does not suffer from
this problem, has been suggested in \cite{GPVW}. 

In the rest of this paper, we shall assume that all pockets are
identical, so the prior is given by Eq.(\ref{PZ}) for symmetric
potentials and by Eq.(\ref{Pq}) with $P_q(X)$ from Eq.(\ref{ergodic})
in the more general case. We note that for power-law potentials like
(\ref{V}), inclusion of the factor $\exp(S)$ does not change the
qualitative character of the result. For such potentials, Eq.(\ref{Z})
gives
\beq
Z\approx \exp\left({4\pi\over{n}}\phi_{qX}^2\right),
\label{Z'}
\eeq
where we have assumed that $\phi_{qX}\gg\phi_{*X}$ (which is always
satisfied for $\lambda\ll 1$). On the other hand, it follows
from Eq.~(\ref{phiq}) that $H(\phi_{qX},X)\sim 1/\phi_{qX}$, and thus
\beq
S(X)\sim \phi_{qX}^2.
\eeq
Hence, the exponents in the volume factor (\ref{PZ}) and in the ergodic factor 
(\ref{ergodic}) are both proportional
to $\phi_{qX}^2$, differing only by a numerical coefficient.

It also follows from Eqs.~(\ref{phiq}),(\ref{V}) that $\phi_{q}\sim
\lambda^{-1/(n+2)}$. Hence, we can write
\beq
P_{prior}(\lambda)\propto \exp(C\lambda^{-2/(n+2)}),
\label{Plambda}
\eeq
where $C\sim 1$ is a constant.

\section{The small-$Q$ catastrophe}

Let us now consider what Eq.~(\ref{Plambda}) implies for the prior
distribution $P_{prior}(Q)$. Substituting the relation (\ref{slambda})
between $Q$ and $\lambda$ in (\ref{Plambda}), we obtain 
\beq
P_{prior}(Q)\propto \exp (C' Q^{-4/(n+2)})
\label{PQ}
\eeq
with $C'\sim 1$. Assuming that $\lambda$ is bounded from above by
$\lambda\lesssim 1$, this equation applies for $Q\lesssim 1$.

The distribution (\ref{PQ}) has a very sharp peak at small $Q$.
The exponential decay towards larger $Q$ overrides any
power-law prefactors, and thus the large $Q$ catastrophe is
averted. But now there is an even bigger problem. According to the
distribution (\ref{PQ}), small values of $Q\sim 10^{-6}$ at the low
end of the anthropic range are many orders of magnitude more probable
than the observed value of $Q\sim 10^{-5}$. Thus, we have a full-blown
{\it small-$Q$ catastrophe}.

This situation is not specific to power-law potentials. Both the
inflationary expansion factor and the ergodic factor depend
exponentially on the parameters specifying $H(\phi)$, so the
exponential form of $P_{prior}(Q)$ is generic.

\section{Curvaton-type models}

The conclusion appears to be that the density perturbations due to 
fluctuations of the inflaton are negligible, and thus the observed
perturbations must have been generated by some other mechanism. This
conclusion has been reached in \cite{AV95}, where it was argued that
the fluctuations are probably due to topological defects. However, CMB
observations have ruled out defects as the dominant source of
perturbations. 

Another mechanism, which is consistent with the data, is the curvaton
model \cite{LM97,LW02}. In the simplest version, it
involves a light scalar field $\chi$ (the curvaton) of mass $m\ll
H$. Quantum fluctuations during inflation randomize the values of
$\chi$, with a Gaussian distribution of width $\sim H^2/m$. This
distribution is nearly flat, $dP(\chi)\propto d\chi$, in the range
$|\chi| \ll H^2/m$. The energy density of the curvaton is negligible
during inflation, but at a later stage it comes to dominate the
universe. The curvaton density perturbations are of the order
\beq
Q \sim \delta\chi/\chi,
\label{dchi}
\eeq
where $\delta\chi\sim H$ is the quantum fluctuation of $\chi$ during
inflation, at the epoch when the relevant scale crossed the
horizon. Eventually, the curvaton decays, and the density perturbation
(\ref{dchi}) is transferred to the radiation.

An alternative class of models \cite{DGZ,Kofman} assumes that $\chi$
is a modulus, whose value determines the masses and/or couplings of
some other particles. The resulting perturbation amplitude is still
given by (\ref{dchi}). In both cases, we can write
\beq
Q\sim H/\chi.
\label{Qchi}
\eeq

To evaluate the prior distribution for $Q$, we shall consider the
simplest model, where the inflaton potential is fixed. If, for
example, the potential is given by Eq.(\ref{V}), then we assume that
$\lambda$ has the same value everywhere in the universe. The
variation of $Q$ is then entirely due to the long-wavelength
fluctuations of the field $\chi$.

The crucial difference from the case of inflaton-induced perturbations
is that the energy of $\chi$ is subdominant during inflation, and thus
the inflationary expansion factor $Z$ is not correlated with $Q$. The prior
distribution for $Q$ is 
\beq
dP_{prior}(Q)\propto d\chi \propto dQ/Q^2,
\label{Qprior}
\eeq
and Eq.(\ref{Ps}) gives
\beq
dP(Q)\propto QdQ.
\label{curvaton}
\eeq

This appears to be an improvement over the models that we considered
so far. The exponential dependence on $Q$ has disappeared, and the
growth of the probability towards larger $Q$ is considerably milder
than before. However, the value of $Q\sim 10^{-4}$ is still 100 times
more probable than the value we observe. We shall now discuss some
ways to address this milder version of the large-$Q$ catastrophe.

\section{Evading the large-$Q$ catastrophe}

A possible attitude is simply to accept that the observed value of $Q$
has probability of about 1\%. This is somewhat uncomfortable, but
perhaps not unreasonably small.

Alternatively, we can consider a multi-component curvaton model with
the potential
\beq
V(\chi)={1\over{2}}m^2\chi^2,
\eeq
where
\beq
\chi^2\equiv \sum_{a=1}^N \chi_a^2.
\eeq
In this case, the distributions (\ref{Qprior}) and (\ref{curvaton})
are replaced by 
\beq
dP_{prior}(Q)\propto \chi^{N-1}d\chi \propto Q^{-(N+1)}dQ
\eeq
and
\beq
dP(Q)\propto Q^{2-N}dQ.
\eeq
For $N=3$, the distribution is flat on the logarithmic scale.

Finally, suppose the distribution indeed rises towards larger
$Q$. Then we expect to have $Q\sim Q_{max}$, where $Q_{max}$ is the
upper boundary of the anthropic range, while the observed value
is $Q\ll Q_{max}$. It is possible, however, that $Q_{max}\sim 10^{-4}$
is an overestimate, and the actual bound is not so far from the
observed value of $Q\sim 10^{-5}$.

The original argument by Tegmark and Rees
\cite{TR} is that, as $Q$ gets larger, galaxies form earlier, so
typical galaxies are smaller and denser. As a result, the rate of
stellar encounters that can disrupt planetary orbits gets larger, and
becomes unacceptably large for $Q\gtrsim 10^{-4}$.

However, it is conceivable that factors other than planetary orbit
disruption play the key role in anthropic selection. One possibility
is that perturbation of comets could be such a factor
\cite{GV03}. Comets move around the Sun, forming the Oort cloud of
radius $\sim 0.1$ pc. Whenever the cloud is significantly perturbed by
a passing star, a rain of comets is sent to the interior of the Solar
system. Occasionally the comets hit the Earth, causing mass
extinctions. Another threat is posed by nearby supernova explosions
(see, e.g., \cite{extinctions} and references therein). Without
attempting to estimate the rate of mass extinctions due to these
effects, we note that the time it took to evolve intelligent beings
after the last major extinction is comparable to the typical time
interval between extinctions ($\sim 10^8$ years). This suggests that
we are indeed close to the boundary of the anthropic range. A
substantial increase in the rate of extinctions might interfere with
the evolution of observers.

\section{Conclusions}

We studied the probability distribution for the cosmological constant
$\Lambda$ and the density contrast $Q$ in models where both of these
parameters are variable. With only mild assumptions about the prior
probabilities, the distribution $P(\Lambda,Q)$ factorizes into two
independent distributions for the variables $Q$ and $y=\Lambda/Q^3
\rho_m$. (Here, $\rho_m$ is the density of nonrelativistic matter.)
The distribution for $y$ is rather model-independent and is in a good
agreement with the observed value of $y$. Thus, despite recent claims
to the contrary, the successful anthropic prediction for $y$ is not
destroyed by allowing $Q$ to vary.

The form of the probability distribution $P(Q)$ depends on the assumed
model for the generation of density perturbations and on the mechanism
responsible for the variation of $Q$ from one part of the universe to
another. We first considered models where the perturbations are due to
quantum fluctuations of the inflaton field $\phi$, and their magnitude
varies due to the variation of the inflaton self-coupling
$\lambda$. (The variation of $\lambda$ may be induced by fluctuations
of some other light field. Alternatively, $\lambda$ may take a dense
discretuum of values in different vacua of the lanscape.) In this
case, $P(Q)$ tends to have an exponential dependence on $Q$. The
reason is that the same inflaton coupling $\lambda$ determines both
the magnitude of $Q$ and the amount of inflationary expansion. The
probability $P(Q)$ is proportional to the expansion factor, which
depends exponentially on $\lambda$, and therefore on $Q$. For
power-law, ``chaotic'' inflation models, we found that $P(Q)$ has an
extremely sharp peak at small $Q$. Models of this kind cannot account
for the observed density perturbations.

We next considered curvaton-type models, where the density
perturbations are generated by one or several light fields $\chi_a$,
other than the inflaton. In this case, the magnitude of $Q$ varies due
to the fluctuations in the long-wavelength component of $\chi_a$, so
there is no need to assume any variation of the couplings. The crucial
difference from the case of inflaton-induced perturbations is that the
energy of $\chi_a$ is subdominant during inflation, and thus the
inflationary expansion factor is not correlated with $Q$. We find
that, as a result, the distribution $P(Q)$ has a rather mild
dependence on $Q$ and can even be {\em flat} in some models, in which
case there is no conflict at all with observations.  Both large and
small-$Q$ catastrophes can thus be averted.

The models we have discussed are the simplest ones, with only two
variable parameters. On the other hand, in the context of string
theory landscape, we expect to have a multitude of variables, and
perhaps none or very few parameters fixed. This multi-parameter space
is awaiting to be explored. Here, we shall only briefly comment on
some issues relevant to the variation of $Q$.

A simple extension of the parameter space is achieved by combining the
two models that we have discussed: the inflaton potential has a
variable coupling $\lambda$, and there are, in addition, some light
curvaton fields. As before, the probability distribution will grow
exponentially towards small $\lambda$. This growth will eventually be
cut off, due to constraints related to thermalization and
baryogenesis.  The resulting distribution will have a sharp peak at a
value of $\lambda \ll 10^{-14}$, so most observers will find $\lambda$
to be close to that value. The density perturbations due to the
inflaton will therefore be negligible. With $\lambda$ now fixed, the
probability distribution for $Q$ will be the same as in curvaton-type
models that we discussed earlier. An extension to more complicated
inflaton potentials, specified by several variable parameters,
requires further investigation.

This work was partially supported by the grants FPA 2004-04582-C02-02
and DURSI 2001-SGR-0061 (J.G.) and by the National Science Foundation
(A.V.).

\end{document}